# Thermal Conductivity above 2000 W/m·K in Boron Arsenide by Nanosecond Transducer-less Time-Domain Thermoreflectance

Hong Zhong[1†], Ying Peng[2†], Feng Lin[1†], Ange Benise Niyikiza[2], Fengjiao Pan[2], Chengzhen Qin[3], Jinghong Chen[1], Viktor G. Hadjiev[4], Liangzi Deng[1], Zhifeng Ren[2*], Jiming Bao[1,2,3*]

[1]Department of Electrical & Computer Engineering and Texas Center for Superconductivity at the University of Houston (TcSUH), University of Houston, Houston, Texas 77204, USA.
[2]Department of Physics and Texas Center for Superconductivity at the University of Houston (TcSUH), University of Houston, Houston, Texas 77204, USA.
[3]Materials Science and Engineering, University of Houston, Houston, Texas 77204, USA.
[4]Department of Mechanical Engineering and Texas Center for Superconductivity at the University of Houston (TcSUH), University of Houston, Houston, Texas 77204, USA.

[†]These authors contributed to the work equally
[*]Corresponding authors. Email: zren@uh.edu, jbao@uh.edu

**Abstract**

Cubic boron arsenide (c-BAs) has been theoretically predicted to exhibit thermal conductivity $\kappa$ comparable to that of diamond, yet experimental measurements have plateaued at ~1300 W/m·K. We report room-temperature $\kappa$ exceeding 2000 W/m·K in c-BAs, on par with single-crystal diamond. This finding is enabled by high-quality single crystals and a newly developed nanosecond, transducer-less time-domain thermoreflectance technique that allows spatial mapping of $\kappa$ without metal transducers. Thermal conductivity correlates with crystal quality, as evidenced by stronger photoluminescence and longer photoluminescence lifetimes. However, the observed nanosecond lifetimes remain shorter than expected for an indirect bandgap semiconductor, suggesting room for further crystal quality improvement and higher $\kappa$. These results challenge current theoretical models and highlight c-BAs as a promising material for next-generation electronics.

Cubic boron arsenide (c-BAs) has established itself as a superior semiconductor that possesses both exceptionally high thermal conductivity and high charge carrier mobility (*1-7*), however, several technical challenges must be addressed before it can be widely adopted for its promising applications. The primary challenge lies in the non-uniform crystal quality in terms of impurity and doping (*6, 7*). This issue was first identified through the observation of highly non-uniform photoluminescence (PL) in seemingly single crystals (*8*). A related hurdle is the lack of a rapid, transducer-less method for measuring thermal conductivity. While traditional time-domain thermoreflectance (TDTR) has been instrumental in discovering high thermal conductivity of c-BAs (*2-4*), it is not widely accessible to the materials research community due to its reliance on an expensive femtosecond laser system and highly specialized data acquisition and analysis techniques (*9, 10*). Additionally, conventional TDTR is an inconvenient technique that requires depositing a metal film as both a light absorber and thermal transducer (*9, 10*). This not only increases measurement complexity and uncertainty but also restricts further materials characterizations and device testing. In addition, the traditional TDTR is typically operated in frequency domain using high frequency modulation and lock-in detection, which makes TDTR slow and difficult to extract thermal conductivity in real-time (*9, 10*).

Efforts have been made to eliminate the metal transducer and develop a transducer-less version of traditional TDTR, but these attempts have faced significant theoretical and practical challenges and have not been very successful (*11-13*). A major issue, recognized early on, is that the pump pulse-induced transient reflectance of the probe pulse is influenced not only by temperature change but also by contributions from photo-excited charge carriers (*11-13*). In metals such as Al or Au, charge carriers excited by the pump have extremely short lifetimes—on the order of femtoseconds—allowing them to rapidly relax and convert into heat. As a result, changes in probe reflectivity are directly related to temperature variations in the metal transducer, *i.e.*, thermoreflectance. In contrast, charge carriers in semiconductors have longer lifetimes, they are excited instantaneously by the pump and stay there after the passing of the pump, meaning that ultrafast transient reflectance is dominated by electronic contributions rather than pure thermoreflectance (*11-13*). This is why femtosecond time-resolved reflectance has traditionally been used to probe coherent electronic states and carrier dynamics (*6, 7, 14, 15*). To mitigate electronic contributions in traditional TDTR and obtain a clearer thermoreflectance signal, researchers have employed either long pump-probe delays (> ns) or low modulation frequencies (*7, 11-13*). However, neither approach fully eliminates electronic contribution. While long pump-probe delays have been somewhat effective for measuring the thermal conductivity of Si and Ge (*11-13*), the signal at such long delays weakens significantly. In many cases, it decays to the point of becoming undetectable after femtosecond excitation, reducing the value of ultrafast time resolution (*13*).

In this work, we demonstrate a nanosecond, transducer-less TDTR (tl-TDTR) technique that overcomes key challenges of traditional TDTR. tl-TDTR enables rapid thermal conductivity ($\kappa$) mapping of c-BAs, identifying many crystals or regions with $\kappa$ exceeding the theoretical limit of 1300 W/m·K, with some surpassing 2000 W/m·K. The accuracy of tl-TDTR is validated through $\kappa$ measurements of well-characterized materials as well as self-consistency by producing the same $\kappa$ values with and without an Al film transducer. To investigate the origin of ultrahigh thermal conductivity, we employ photoluminescence (PL) lifetime measurements, revealing that higher-$\kappa$ c-BAs exhibit longer PL lifetimes. However, the relatively short PL lifetime of c-BAs suggests that its ultimate thermal conductivity limit has yet to be reached. The discovery of ultrahigh thermal conductivity not only expands the potential applications of c-BAs but also challenges existing theoretical models, calling for a deeper understanding of the underlying mechanisms.

Figs. 1A and Bi illustrate the experimental setup for tl-TDTR. A 527-nm nanosecond laser (~80 ns, 1 KHz repetition rate) serves as the pump, while a 450-nm continuous-wave (CW) laser functions as the probe. Both beams are focused on the same spot through a microscope objective, with pump and probe spot sizes of ~30 µm and ~5 µm, respectively. The pump power is varied from 0.1 to 2.5 mW depending on the material's reflectivity and absorbance, while the probe power is maintained below 2 mW. To achieve high time resolution and signal gain, we use a fast balanced photodetector to continuously monitor the reflected probe, while a digital oscilloscope rapidly acquires and averages the signal over hundreds of iterations. The high repetition rate of the pump allows for efficient averaging over thousands of cycles, ensuring a strong signal-to-noise ratio without excessive measurement time. By eliminating the mechanical pump-probe delay rail, high-frequency pump modulation, and lock-in detection of the reflected probe, tl-TDTR enables real-time acquisition of thermoreflectance traces. Multi-physics COMSOL modeling is employed to simulate optical heating and subsequent heat transfer dynamics (Fig. 1Ci), it is flexible to accommodate various configurations, including different pump spot sizes and the presence or absence of a metal film transducer (see Supplementary Material for details). Key modeling parameters, such as the optical absorption coefficient and thermoreflectance coefficient are obtained from literature or measured in the lab. The pump beam is modeled as a Gaussian beam (Figs. S1 and S2 in Supplementary Material), with diameter and power values derived from experimental measurements.

Figs. 1D-F show representative tl-TDTR traces for Si, InP, and Ge. For Si, the relative change in reflectance ($\Delta R/R$) has been converted to a temperature change $\Delta T = \Delta R/(R \cdot C_{TR})$ using its thermoreflectance coefficient $C_{TR} = 1.3 \times 10^{-4}$ K$^{-1}$ (*16*). Each trace exhibits a sharp reflectance peak immediately after pump pulse excitation, followed by a gradual decrease, indicating rapid heating by the pump pulse and subsequent cooling. These dynamics closely resemble traditional TDTR behavior, albeit on a different timescale, confirming that tl-TDTR signals are primarily governed by thermoreflectance (*7, 11-13*). The width of the initial peak is largely determined by the pump pulse duration, while the subsequent cooling rate reflects the material's thermal conductivity. The experimental curves are well-fitted using COMSOL simulations (dotted lines), yielding κ values of 150, 68, and 58 W/m·K for Si, InP, and Ge, respectively—fairly consistent with reported values (*12, 13, 17*). It is important to note that the actual value of the thermoreflectance coefficient will not affect the heat transport dynamics; rather, it influences the sensitivity of temperature measurement, similar to the role of the metal film in traditional TDTR. Therefore, we normalize the $\Delta R/R$ traces during COMSOL fitting and are able to extract the thermal conductivities of Ge and InP without knowing their thermoreflectance coefficients.

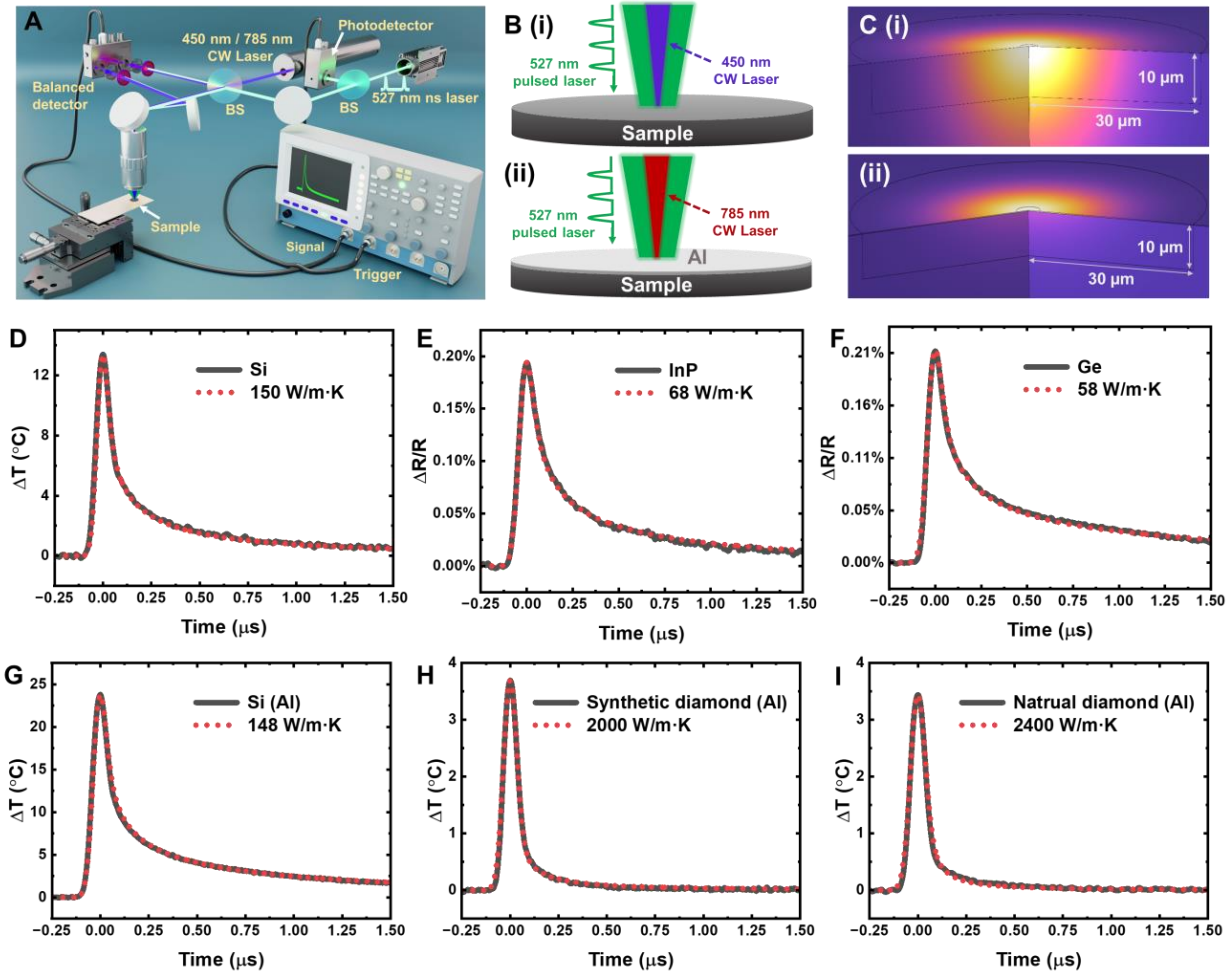

**Fig. 1. Demonstration of nanosecond TDTR with well-characterized materials, both without and with an Al film as a transducer.** (A and Bi) Schematics of the experimental setup for nanosecond transducer-less time-domain thermoreflectance (tl-TDTR). (Ci) COMSOL simulation model (actual simulation domain: 300 µm tall and 500 µm in radius). (D-F) Representative tl-TDTR traces of Si, InP, Ge (black solid lines) and corresponding COMSOL fits (red dotted lines). (A, Bii, Cii) Schematics and COMSOL simulation model for tl-TDTR with an Al transducer. (G-I) Representative tl-TDTR traces with Al coating for Si, synthetic diamond, natural single crystal diamond (black solid lines) and their COMSOL fits (red dotted lines).

Note that in previous transducer-less TDTR experiments using a femtosecond pump laser, the traces were dominated by strong electronic contributions, evident as a pronounced negative transient reflectance (*11-13*). This issue is unavoidable in semiconductors because the lifetime of photoexcited carriers exceeds the femtosecond pulse width. These challenges almost vanish with our tl-TDTR setup because of the following reasons. First, there is a fundamental difference between carrier and thermal diffusion. The diffusion and population of the carriers are limited by their lifetime; For instance, under steady-state photoexcitation, the carrier population remains constant and confined within a fixed diffusion length. In contrast, the temperature of photoexcited region continuously rises as thermal energy accumulates and spreads over greater distances. By increasing the pulse duration from ~100 femtoseconds to ~100 nanoseconds—a six-order-of-magnitude change, the thermal contribution completely overwhelms the electronic signal. Furthermore, under pulsed laser excitation, the lifetime of photoexcited carriers is significantly reduced compared to CW excitation due to Auger and two-body recombination processes (*12*). A third contributing factor is our choice of probe wavelength. The charge carrier-induced reflectivity decreases significantly at shorter probe wavelengths (*6, 18*). In our previous work, for example, the reflectivity dropped by a half when the probe wavelength was changed from 580 nm to 530 nm (*6*). Finally, it is important to note that a significant amount of heat is generated almost instantaneously, as in metals, when high-energy photoexcited electrons relax to the conduction band edge, with thermal energy continuing to accumulate during photoexcitation. Together, these designs of TDTR ensure that tl-TDTR exclusively measures thermoreflectance.

The tl-TDTR also works with a thin metal transducer—which is necessary when a semiconductor's bandgap exceeds the pump laser energy. To demonstrate this flexibility and further confirm the accuracy of tl-TDTR, we chose Si, synthetic diamond, and natural single-crystal diamond for validation. We deposited a thin Al layer on them and selected a 785 nm probe laser because Al exhibits a larger thermoreflectance temperature coefficient at that wavelength (Figs. 1A and Bii). This thin Al film can be modeled in COMSOL by introducing an interfacial thermal conductance as a fitting parameter (Fig. 1Cii). Figs. 1G-I show new tl-TDTR traces for Si and diamond along with COMSOL fits for all three samples. We obtained the same $\kappa$ for silicon and high values of 2000 and 2400 W/m·K for synthetic and natural diamonds, respectively, in agreement with accepted values (*19*). We continue to use the term tl-TDTR even when an Al transducer is employed, as the underlying technique remains the same. To avoid confusion, no transducer will be used unless otherwise specified. These results further validate the accuracy of tl-TDTR across a wide range of thermal conductivities and demonstrate its versatility for characterizing both thin films and wide-bandgap semiconductors.

One advantage of tl-TDTR is that relative thermal conductivity can be quickly assessed by comparing cooling rates—faster cooling indicates higher conductivity. Given the non-uniformity of c-BAs, we first scanned the cooling rate using tl-TDTR and then selected specific regions for detailed COMSOL fitting. Fig. 2A presents a scanning electron microscopy (SEM) image of a new c-BAs sample, while Fig. 2B displays a manually acquired cooling rate map, defined as the time from the peak to half its value. We selected three representative spots (P1, P2, and P3) for further analysis. Figs. 2C-E show their tl-TDTR traces with COMSOL fittings, revealing that spot P1 exhibits a thermal conductivity of 1500 W/m·K – higher than the theoretical prediction of 1300 W/m·K. Here the temperature change $\Delta T$ is calculated based on the measured thermoreflectance coefficient, which is approximately $3 \times 10^{-4}\,\text{K}^{-1}$ at 450 nm. To verify this high thermal conductivity, we evaporated 100 nm of Al, switched the probe to a 785-nm laser, and measured thermal conductivity at the same locations. Figs. 2F–H display the corresponding TDTR traces and COMSOL fittings, confirming that all three spots yielded the same thermal conductivity values.

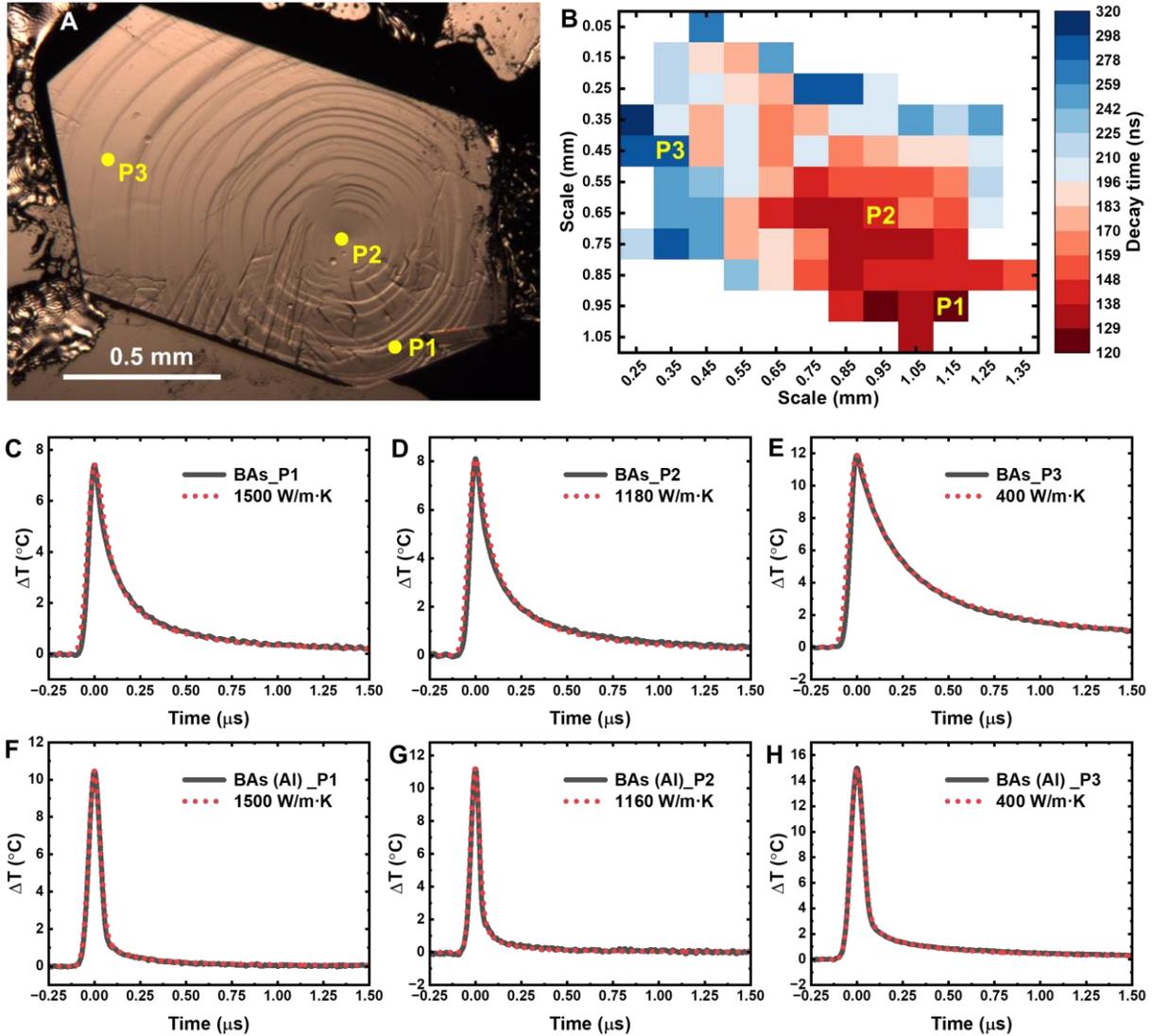

**Fig. 2. Nanosecond TDTR measurements of c-BAs with and without an Al transducer.** (A) Scanning electron microscopy (SEM) image. (B) tl-TDTR decay time map. (C-E) tl-TDTR traces (black solid lines) and corresponding COMSOL fits (red dotted lines) for three spots with thermal conductivities of 1500, ~1180, and 400 W/m·K, respectively. (F-H) TDTR traces (black solid lines) and COMSOL fits (red dotted lines) with Al transducer for the same three spots.

Fig. 2 also reveals that TL-TDTR traces with and without an Al transducer exhibit markedly different heat transport dynamics. With the Al transducer, the thermoreflectance peak closely follows the pump pulse, producing a sharp response. The temperature drops rapidly after the peak, whereas in tl-TDTR, the temperature decreases much more gradually. This difference arises from the distinct heating mechanisms by the pump laser pulse. tl-TDTR involves volumetric heating (*11*), as the pump pulse penetrates to a depth of ~14 µm, based on UV-Vis absorbance data (*8*). In contrast, TDTR with a metal transducer involves surface heating due to the extremely shallow optical penetration depth. In fact, in traditional TDTR modeling, the metal transducer is often treated as a mathematically infinitesimal. As a result, heat rapidly accumulates at the metal film and dissipates through interfacial thermal conductance. This contrast is clearly illustrated in the COMSOL simulations in Figs. 3A-B, where the temperature of the Al film is uniform and significantly higher than that of the c-BAs crystal during the heating by pump pulse. In both cases, the rate of temperature decay reflects the underlying thermal conductivity of the materials. However, because the cooling of the Al transducer is also influenced by interfacial thermal conductivity, tl-TDTR provides a more direct measurement of the BAs crystal's thermal conductivity.

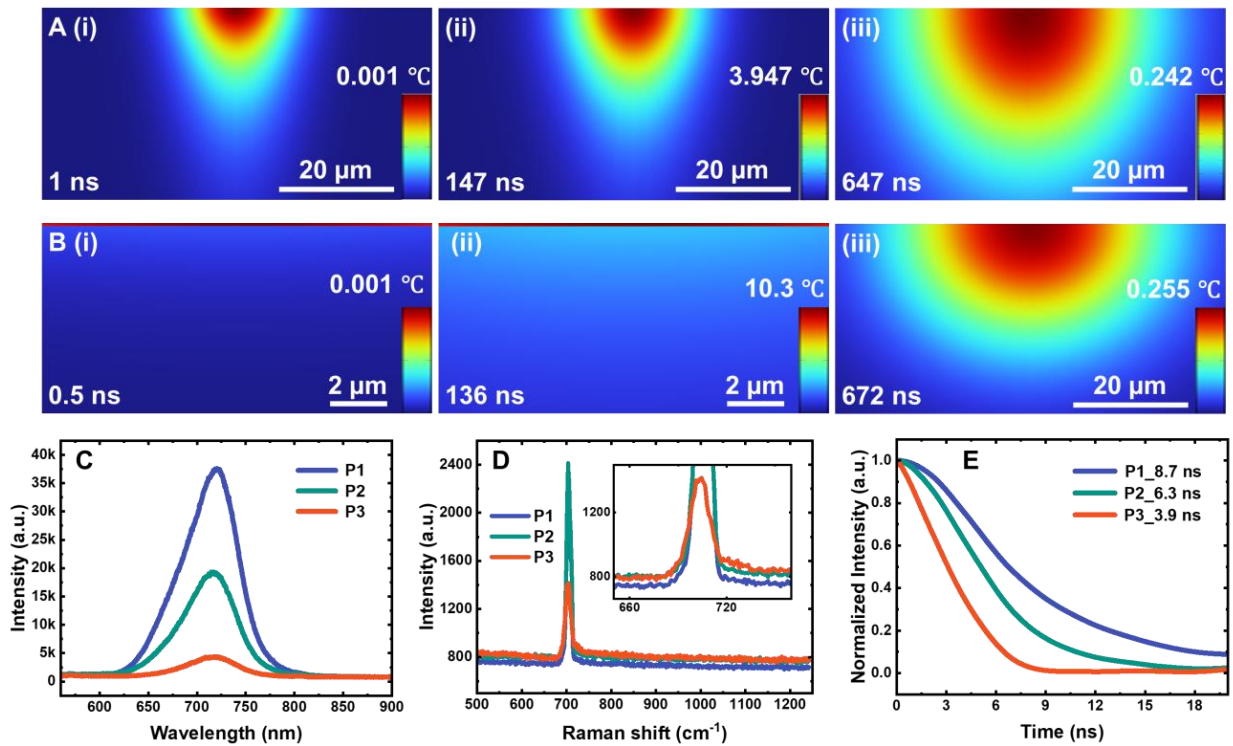

**Fig. 3**. **Comparison of nanosecond TDTR simulations with and without an Al transducer, and optical characterization of c-BAs**. (A-B) COMSOL simulations for spot P1 of c-BAs without (A) and with (B) an Al transducer. (C-E) Photoluminescence (PL), Raman and PL lifetime measurements for spots P1, P2 and P3.

The significantly higher thermal conductivity of P1 compared to P2 and P3 suggests that P1 has a lower level of impurities or defects. To verify this, we measured and compared their PL and Raman spectra (performed before Al evaporation), both of which are commonly used to identify regions with high thermal conductivity. Fig. 3C presents the PL spectra of the three spots. P1 exhibits the strongest PL intensity, followed by P2, while P3 has the weakest. All three spectra are dominated by the bandgap PL peak centered at 720 nm, indicating the overall high quality of the sample (*8, 20*). Fig. 3D shows the Raman spectra from 200 to 1200 cm$^{-1}$. P1 displays the lowest Raman background and the weakest Fano lineshape of the LO phonon at 700 cm$^{-1}$, further supporting its superior crystal quality (*3, 7*). Based on the PL and Raman data, it is evident that P1 has the fewest impurities or defects. This higher crystal quality is what enables P1 to achieve the highest thermal conductivity. To obtain a better understanding of the crystal quality of P1, we measured the PL lifetime - an established parameter for assessing silicon quality but not yet reported for c-BAs (*21*). A longer PL lifetime indicates higher crystal quality, characterized by fewer defects or impurities and reduced non-radiative recombination. For PL lifetime measurements in c-BAs, we used a nanosecond laser (3-5 ns, 532 nm) to excite the sample and a fast photomultiplier tube (PMT) to detect PL signals. The time-resolved PL traces of the three spots are shown in Fig. 3E: P1 exhibits the longest PL lifetime, while P3 has the shortest. Overall, these nanosecond-scale lifetimes are comparable to those of direct bandgap semiconductors but are significantly shorter than those of high-quality indirect bandgap semiconductors such as Si and GaP (*21, 22*). Given its longer PL lifetime, we conclude that P1 is expected to exhibit higher thermal conductivity than P2 and P3.

To further validate this ultrahigh thermal conductivity, we surveyed several additional crystals using the rapid tl-TDTR technique. Multiple spots with thermal conductivity exceeding 1500 W/m·K were quickly identified, with some surpassing this value even further. Fig. 4A shows an optical image of a high-quality sample, Figs. 4B-C display normalized tl-TDTR traces from three spots exhibiting high thermal conductivity. They exhibit very fast but different temperature cooling speeds. Figs. 4D-F show their corresponding COMSOL fits, revealing even higher thermal conductivities of 1800, 2200, and 2400 W/m·K. To illustrate the uncertainty in thermal conductivity measurements, we also include COMSOL fits with values adjusted by ±5%. As shown, deviations of 5% result in imperfect fits, highlighting the sensitivity and accuracy of our measurements.

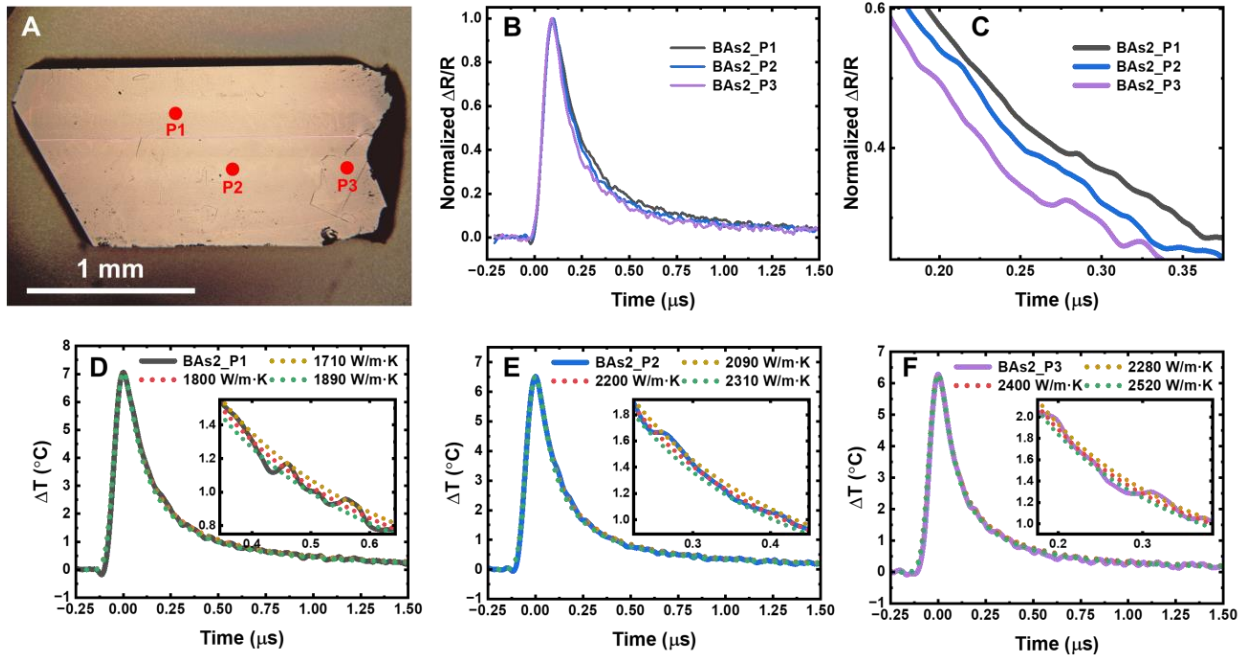

**Fig. 4. c-BAs crystal with thermal conductivity >2000 W/m·K**. (A) Image of a high-quality c-BAs showing three high-κ spots. (B) Normalized tl-TDTR traces for the three spots, illustrating different cooling rates. (C-D) COMSOL fits (dotted lines) for the three tl-TDTR traces. Fits with κ ±5% are also included for comparison.

The observation of thermal conductivity exceeding 2000 W/m·K is unexpected but can still be understood from an experimental perspective. First, the seemingly excellent agreement between experiment and theory is not always consistent - exceptions exist. For example, a recent study measured the thermal conductivity of c-BAs with a silicon impurity concentration of $3 \times 10^{19}$ cm$^{-3}$ to be 920 W/m·K (*23*). However, theoretical predictions suggest it should be less than half of this value (*7, 24*). This discrepancy indicates that theoretical models can significantly underestimate thermal conductivity, suggesting that in ultrapure c-BAs, the actual thermal conductivity could far exceed the theoretical limit of 1300 W/m·K. On the other hand, based on PL lifetime measurements, spot P1 cannot be classified as an ultrapure crystal. In ultrapure c-BAs, the PL lifetime is expected to reach at least the order of hundreds of nanoseconds, as observed in other indirect bandgap semiconductors (*21, 22*).

In conclusion, we have developed a nanosecond, transducer-less TDTR technique capable of rapidly screening thermal conductivity in semiconductors without the need for depositing metal transducers. tl-TDTR can be used to map the thermal conductivity of most semiconductors without depositing a metal film. For wider bandgap semiconductors, a nanosecond UV laser can be used. Using tl-TDTR, we identified c-BAs samples with thermal conductivity exceeding 2000 W/m·K. PL lifetime studies suggest that this higher-than-predicted thermal conductivity stems from improved crystal quality compared to previous samples. However, an even longer PL lifetime is expected, indicating the potential for even higher thermal conductivity. Since thermal conductivity and electron mobility are closely correlated, the observation of ultrahigh thermal conductivity suggests the possibility of achieving ultrahigh electron mobility as well. Determining the new upper limits of thermal conductivity and electron mobility requires further investigation. The discovery of thermal conductivity surpassing theoretical predictions not only necessitates the development of new theory models but also expands the potential applications of c-BAs.